\documentstyle[aps,pra,epsf]{revtex}
\title{Level Dynamics and Universality of Spectral Fluctuations}
\author{Peter Braun$^{1,2}$, Sven Gnutzmann$^{1,5}$, Fritz Haake$^{1}$,
Marek Ku\'s$^{1,3}$, and
Karol \.Zyczkowski$^{1,3,4}$}
\address{
$^{1}$Fachbereich Physik, Universit\"at-GH Essen,
Essen, Germany\\ $^{2}$Institute of Physics,
Saint-Petersburg University, Saint-Petersburg, Russia\\ 
$^{3}$Centrum Fizyki Teoretycznej, Polska Akademia Nauk, 
Warszawa, Polska\\
$^{4}$Instytut Fizyki, Uniwersytet Jagiello{\'n}ski, 
Krak{\'o}w, Polska\\
$^{5}$Department of Physics of Complex Systems, 
The Weizmann Institute of Science, Rehovot, Israel
}

\begin{document}
\maketitle
\begin{abstract}
The spectral fluctuations of quantum (or wave) systems with a chaotic
classical (or ray) limit are mostly universal and faithful to
random-matrix theory. Taking up ideas of Pechukas and Yukawa we
show that equilibrium statistical mechanics for the fictitious
gas of particles associated with the parametric motion of
levels yields spectral fluctuations of the random-matrix type. 
Previously known clues to that goal are an appropriate equilibrium 
ensemble and a certain ergo\-dicity
of level dynamics. We here complete the reasoning by establishing a power 
law for the $\hbar$ dependence of the mean parametric separation of avoided 
level crossings. Due to that law universal spectral fluctuations emerge as 
average beha\-vior of a family of quantum dynamics drawn from a control
parameter 
interval which becomes vanishingly small in the classical limit; the family 
thus corresponds to a single classical system. We also argue that 
classically integrable dyna\-mics cannot produce universal spectral
fluctuations since their level dynamics resembles a nearly
ideal Pechukas-Yukawa gas.
\end{abstract}
\section*{\strut}
Most quantum systems whose classical limit
is chaotic display universal spectral fluctuations
\cite{BGS,Berry,bible,Bibel}. Such universality was first observed in nuclear
spectra, but is by now well supported by atomic and molecular
spectroscopy. Somewhat surprisingly at first came
support from microwave resonators and later
even from elastic vibrations of crystals, i.~e.~classical
waves for which the ray limit is chaotic. Such experimental as
well as numerical findings suggest that discrete spectra of
non-integrable quantum or classical waves display universal fluctuations.

The universal fluctuations in question can be roughly characterized as
level repulsion. They are also met with in spectra of random Hermitian
or unitary matrices drawn from the Gaussian or circular
ensembles of random-matrix theory \cite{Mehta} pioneered by Wigner and
Dyson. They contrast to the spectral fluctuations of integrable
systems with two or more freedoms where we find level clustering, just
as if the levels followed one another like the events in a Poissonian
random process.

The general rule just sketched is not without exceptions, some of which
will concern us below. It is widely accepted, though, that hyperbolic
classical dynamics can fail to being faithful to random-matrix theory
(in their level-spacing distribution as well as low-order correlation
functions of the level density) only due to either quantum or Anderson
localization or to slight perversity, like symmetries relevant for
quantum spectra but not coming with classical constants of the motion
\cite{BogomolnyGeorgeotGiannoniSchmit,KeatingMezzadri}. Faithfulness to
random-matrix theory even prevails for dynamics with mixed phase spaces,
as long as islands of regular motion remain smaller than Planck cells.

Theoretical attempts to explain the applicability of random-matrix
theory to quantum and wave chaos have been based on treating the
parametric dependence of spectra with the methods of statistical
mechanics \cite{Bibel,Yukawa,Wilkinson} and, more recently, on extending ideas
from the theory of disordered systems to dynamical systems
\cite{MIT,Zirnbauer}.
Building on the first of these strategies we here propose to put forth
arguments which could lead to clarifying the issue.

Parametric level dynamics was related by Pechukas \cite{Pechukas} and
Yukawa \cite{Yukawa} to the classical Hamiltonian motion of a
one dimensional gas, the Pechukas-Yukawa
gas (PYG). Different PYG's arise \cite{Bibel} for Hamiltonians
depending on a parameter $\tau$ as $H(\tau)=H_0+\tau H_1$ or
$H(\tau)=H_0\cos\tau+H_1\sin\tau$ or $H(t,\tau)=H_0+\tau
H_1\sum_{i=1,2,\ldots}\delta(t-i)$. The latter involves
periodic kicks; the time evolution over one period is
described by unitary Floquet operators of the form $F={\rm e}^{-{\rm
i}H_0}{\rm e}^{-{\rm i}\tau H_1}$ whose quasienergies $\varphi_m$
are defined by $F|m,\tau\rangle={\rm e}^{-{\rm i}\varphi_m}|m,\tau\rangle$.
Note that we have taken the liberty to absorb a unit of time as well as
Planck's constant in the Hamiltonians so as to make them as well as the
perturbation strength $\tau$ dimensionless.

We find it convenient to base the following arguments on the PYG for
periodically kicked systems (but we shall occasionally draw in
illustrations for autonomous systems as well). The quasienergies
$\varphi_m(\tau)$ with $m=1\ldots N$ appear as the coordinates of
particles changing as the ``time'' $\tau$ elapses. The role of conjugate
momenta is taken up by $p_m(\tau)=\langle m,\tau|H_1|m,\tau\rangle$. The
Hamiltonian function comprises a kinetic term and a pair interaction
\cite{Bibel},
\begin{equation}
   {\cal H}=\frac{1}{2}\sum_{m=1}^Np_m^2 + \frac{1}{8}\sum_{m\neq n}
             \frac{|l_{mn}|^2}{\sin^2\frac{\varphi_m-\varphi_n}{2}}\,.
   \label{1}
\end{equation}
As the only unusual feature of that Hamiltonian we encounter
coupling strengths $l_{mn}$ which are themselves
dynamical variables, related to the
off-diagonal elements of the perturbation as $l_{mn}={\rm i}\langle
m,\tau|H_1|n,\tau\rangle ({\rm e}^{{\rm i}(\varphi_m-\varphi_n)}-1)$. If
time reversal invariance holds for the quantum dynamics
(for the other universality classes similar considerations apply),
the initial $l_{mn}(0)$ can be chosen to form a real antisymmetric matrix
with $N(N-1)/2$ independent elements, $l_{mn}=-l_{nm}=l_{mn}^*$, and the
dynamics of the gas preserves that property. To extract
Hamiltonian equations of motion for the $\varphi_m,p_m,l_{mn}$ we need to
employ the familiar Poisson brackets for canonical pairs
$\{p_m,\varphi_n\}=\delta_{mn}$; the $l_{mn}$, on the other hand, Poisson
commute with the $\varphi_m,p_m$ and between themselves obey
$\{l_{mn},l_{ij}\}=\frac{1}{2}(\delta_{mj}l_{ni}+
\delta_{ni}l_{mj}-\delta_{nj}l_{mi}-\delta_{mi}l_{nj})$.
The latter Poisson brackets happen to be the ones for generators of
rotations in $N$-dimensional real vector spaces. The number of
variables is $D=2N+N(N-1)/2$. Incidentally, if we
were to solve the resulting Hamilton equations through a power series in
the ``time'' $\tau$ we would recover perturbation theory.

Prior to appearing as a concise formulation of parametric level dynamics
or perturbation theory, the PYG Hamiltonian had been known to
mathematicians \cite{CalogeroMoserWojciechowski} and noted for being
integrable, in spite of the non-trivial two-body interaction.
It is easy to see that the PYG moves on an
$N$-torus in its $D$-dimensional phase space: In the eigenrepresentation
of the perturbation, \,$H_1|r\rangle=\omega_r|r\rangle$, the Floquet
matrix takes the form $F_{rs}=({\rm e}^{-{\rm i}H_0})_{rs}{\rm e}^{-{\rm
i}\tau\omega_s}$. The $\tau$-dependence thus lies in the $N$
exponentials ${\rm e}^{-{\rm i}\tau\omega_s}$ which define an
$N$-torus. Going from the $H_1$-representation to the $F$-representation
amounts to a reparametrization of phase space by a canonical
transformation which can deform but not topologically destroy the torus.
If the $N$ eigenvalues $\omega_r$ are incommensurate the $N$-torus will
be covered ergodically as the ``time'' $\tau$ elapses.

When applying equilibrium statistical mechanics to the PYG we must
resist the temptation to employ the canonical ensemble
$\rho(\varphi,p,l)\propto {\rm e}^{-\beta{\cal H}}$ since that ensemble
allows the PYG to visit everywhere on the $(D-1)$-dimensional energy
shell with uniform probability while only the much smaller $N$-torus
mentioned is accessible. It was nevertheless an important observation
\cite{Yukawa} that ${\rm e}^{-\beta{\cal H}}$ gives precisely the
eigenphase density of random-matrix theory, $\rho(\varphi)=\int
d^Np\,d^{N(N-1)/2}l\;{\rm e}^{-\beta{\cal H}}\propto\prod_{m<n}|{\rm
e}^{-{\rm i}\varphi_m}-{\rm e}^{-{\rm i}\varphi_n}|$, as is easily
checked by doing the Gaussian integrals over the $p$'s and $l$'s. The
appropriate ensemble to use for the PYG is the generalized canonical
ensemble
\begin{equation}
  \rho(\varphi,p,l)\propto \exp{\Big\{-\sum\nolimits_{\mu=1}^{D-N}\beta_{\mu}
  C_{\mu}\Big\}}\,;
  \label{3}
\end{equation}
the accessible $N$-torus is thus nailed down by fixing $D-N$ suitable
constants of the motion $C_{\mu}$ in the ensemble mean with the help
of Lagrange parameters $\beta_{\mu}$; one of these $C_{\mu}$
should be the Hamiltonian $\cal H$. That ensemble has been shown
\cite{Babsi} to have the distribution of level spacings as well as
low-order correlation functions of the level density in common with
random-matrix theory, to within corrections of order $1/N$. A good step
towards explaining the universality of spectral fluctuations was thus
taken, but several more steps are still ahead of us.

\begin{figure}[htb]
\begin{center}
\leavevmode \epsfxsize=0.55\textwidth \epsffile{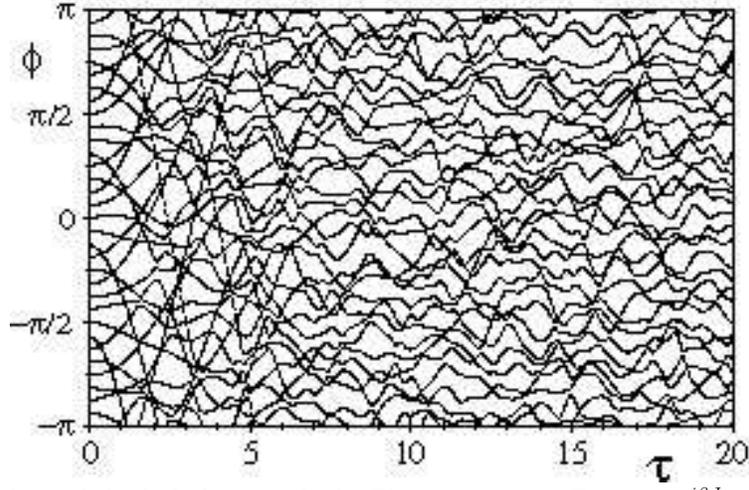}
\caption{Quasienergy levels $\varphi$ of the kicked top with the
Floquet operator $F(\tau)=e^{-i\beta J_{y}}e^{-i\alpha J_{x}}
e^{-i\frac{\tau}{2j+1}J_{x^{2}}}$ for $j=15$, $\alpha=0.907$, 
$\beta=1.072$. As $\tau$ grows from 0 to $\sim 5$ the spectrum ``relaxes'' 
from level clustering to level repulsion, while the dynamics goes from 
integrable to chaotic.}
\label{fig.1} 
\end{center}
\end{figure}

As the phase-space trajectory of the fictitious gas winds around the
$N$-torus, the original Floquet matrix $F(\tau)$ changes within a
one-parameter family. It is that family rather than a single dynamical
system (which has a fixed value of $\tau$) which is assigned random-matrix
type spectral fluctuations. Indeed, ergodicity on the $N$-torus implies
that $\tau$ averages of spectral characteristics like the spacing
distribution $P(S,\tau)$ equal ensemble averages, but the instantaneous
form of $P(S,\tau)$ is thus not specified. To improve the status of
equilibrium statistical mechanics we inquire about the
minimal time window $\Delta \tau$ needed for time and
ensemble averages to become equal. One might demand $\Delta\tau
\to\infty$. Such generous a provision would allow to accommodate and
render weightless equilibration processes originating from initial
conditons far out of equilibrium. As illustrated in Fig.~\ref{fig.1},
equilibration does in fact take place in the spectrum of $F(\tau)={\rm
e}^{-{\rm i}H_0}{\rm e}^{-{\rm i}\tau H_1}$ if $H_0$ is integrable and
chaos develops only as the perturbation is switched on; beyond a certain
``relaxation time'' the phase space regions of regular motion have
shrunk to relatively negligible weight. If, on the other hand, $H_0$ and
$H_1$ are both non-integrable (and from the same universality class) the
initial state of the fictitious gas is already close to equilibrium, and
then the window $\Delta\tau$ in question need not be much larger than
the collision time $\tau_{\rm coll}$ of the gas, i.~e.~the mean distance
of avoided crossings for a pair of neighboring levels.

We now propose to argue that $\tau_{\rm coll}$ scales with the number of
levels $N$ and thus, due to Weyl's law, with Planck's constant like a
power,
\begin{equation}
  \tau_{\rm coll}\propto N^{-\nu}\propto\hbar^{f\nu} \,,\quad\nu>0,
  \label{4}
\end{equation}
where $f$ is the number of degrees of freedom. A $\tau$-average for the
PYG over a window $\Delta\tau\propto\tau_{\rm coll}$ thus involves a
family of Floquet operators $F(\tau)$ which all yield identical
classical dynamics in the limit $\hbar\propto N^{-\frac{1}{f}}\to0$. 
That insight grew out of an idea put forward in Ref.\cite{Zirnbauer}.

The following estimates support the power law (\ref{4}) and reveal
the exponent $\nu$ as non-universal. First, consider a
Hamiltonian $H(\tau)=H_0\cos\tau+H_1\sin\tau\approx H_0+\tau H_1$ and
let $H_0$ and $H_1$ be independent random $N\times N$ matrices drawn
from the Gaussian orthogonal ensemble. Both should have zero mean,
$\overline{H_{0ij}}=\overline{H_{1ij}}=0$, and the same variance,
$\overline{|H_{0ij}|^2}=\overline{|H_{1ij}|^2}=1/N$. The mean level
spacing then is $\Delta\equiv \overline{E_{i+1}-E_i}={\cal
O}(\frac{1}{N})$. The level velocity vanishes in the ensemble mean,
$\overline{H_{1ii}}=0$, and has the variance $p^2\equiv
\overline{H_{1ii}^2}={\cal O}(\frac{1}{N})$. The
estimate $\tau_{\rm coll}\approx \Delta/p\propto N^{-\frac{1}{2}}$ thus yields
the exponent $\nu=\frac{1}{2}$. The same value results for Floquet
operators $F(\tau)={\rm e}^{-{\rm i}H_0}{\rm e}^{-{\rm i}\tau H_1}$
with $H_0,H_1$ random as before. 

Another example is provided by the kicked top \cite{Bibel} 
of Fig.~\ref{fig.1} for which the
operators $H_0$ and $H_1$ are functions of an angular momentum $\vec{J}$
such that $\vec{J}^2=j(j+1)$ is conserved; for fixed $j$ the
dimension of the Floquet matrix is $N=2j+1$. If we choose the
perturbation as a torsion, $H_1=J_z^2/(2j+1)$, the mean level velocity
is $\overline{v}\equiv \frac{1}{N}\sum_{i=1}^NH_{1ii}=\frac{1}{N}{\rm
Tr}H_1=\frac{N}{12}(1+{\cal O}(\frac{1}{N}))$; this gives a common drift
of all levels irrelevant for collisions. A typical relative
velocity $p$ may be found from the variance
$p^2=\frac{1}{N}\sum_{i=1}^N(H_{1ii}-\overline{v})^2$, where the matrix
elements are meant in the eigenrepresentation of
$F(\tau)$; but in that representation the perturbation
$\tilde{H_1}=H_1-\overline{v}$ will, for chaotic dynamics, look like a
full matrix with $\sum_i \tilde{H}_{1ii}\approx 0$ and 
$\sum_i \tilde{H}_{1ii}^2\approx
\frac{1}{N}\sum_{ij}\tilde{H}_{ij}^2=\frac{1}{N}{\rm
Tr}\tilde{H}_{1}^2=\frac{N^2}{180}(1+{\cal O}(\frac{1}{N}))$; we conclude
$p\propto N^{\frac{1}{2}}$. The mean level spacing $\frac{2\pi}{N}$
yields the collision time $\propto N^{-\frac{3}{2}}$ and the exponent
$\nu=\frac{3}{2}$; we have confirmed the latter value by numerically
following level dynamics for $10<j<160$.

\begin{figure}[htb]
\begin{center}
\leavevmode \epsfxsize=0.5\textwidth \epsffile{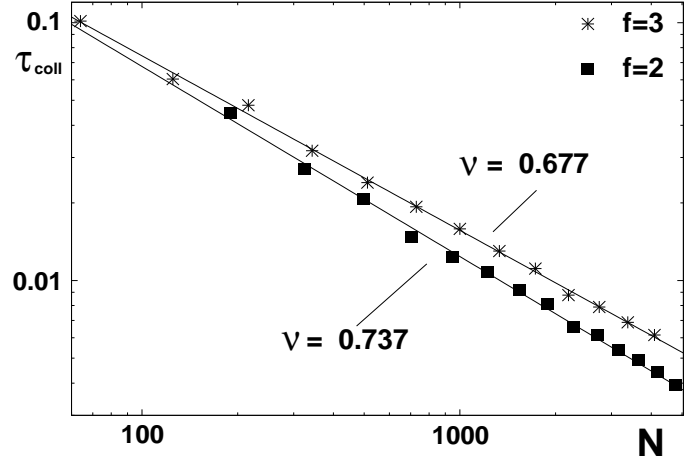}
\caption{$N$-dependence of the level collision ``time'' $\propto N^{-\nu}$ for a Hamiltonian 
      $H(\tau)=H_0\cos\tau+H_1\sin\tau$ with
      $H_{0}, H_{1}$ polynomials in $SU(3)$ generators. Such an $H(\tau)$ can
      correspond to either 2 or 3 classical freedoms.}
\label{fig.2}
\end{center}
\end{figure}

Somewhat daring but interesting are the following estimates for
some classes of Hamiltonians with $f$ freedoms. It is not untypical to
find $H$ represented by a sparse banded $N\times N$ matrix 
with a zebra-like structure of non-vanishing elements. 
An example for such a Hamiltonian is given by a finite cubic lattice
with $N=M^{f}$ sites and nearest and next-to-nearest neighbor interactions.
Non-vanishing matrix elements then only exist in approximate distances
$M^{0},M^{1},\dots,M^{f-1}$ from the diagonal. The overall bandwidth is thus
$M^{f-1}=N^{(f-1)/f}$. 
For similar Hamiltonians we have 
numerically studied the collision time,
choosing
$H(\tau)=H_0\cos\tau+H_1\sin\tau$
as functions of $SU(3)$
generators, like the Lipkin-Meshkov-Glick Hamiltonian known from the
nuclear shell model \cite{Lipkin}. Such $SU(3)$ dynamics can also arise
for collections of identical three-level atoms collectively interacting
with (nearly) resonant modes of a microwave or optical cavity
\cite{Gnutzmann}, and such realizations could even access the
(semi)classical limit as the number of atoms grows large. That limit
can, for a given Hamiltonian, lead to classical motion with either $f=2$
or $f=3$, depending on the representation the initial state
belongs to. We have chosen a Hamiltonian yielding global chaos in both
cases and numerically followed level dynamics over two decades of $N$
(see Fig.~\ref{fig.2}).
The collision time exponents $\nu$ were measured
using $\tau_{\rm coll}=\Delta/p$ and were
found as $0.737$ in the two-freedom
case and $0.677$ for $f=3$.
These numerical results lead us to conjecture
that the collision time for this class of  Hamiltonians
has the exponent $\nu_f=\frac{f+1}{2f}$. The
limit $f\gg1$ corresponds to the above estimate $\nu=\frac{1}{2}$ for
full random matrices.

We now turn to integrable dynamics and their PYG's. The Hamiltonian (\ref{1})
still applies. Only the initial conditions for ${\phi,p,l}$, i.e. the matrices
$H_0,H_1$, tell the PYG that the classical variant of $F(\tau)$ is
integrable. 
We propose to show that the PYG then tends to behave like a near ideal gas. 
Assuming $f$
freedoms and the Hamiltonian expressed in terms of action variables
$I_1,\ldots, I_f$ the semiclassical Einstein-Brillouin-Keller (EBK)
approximation for the spectrum results from letting each action run
through integer or half integer multiples of Planck's constant,
$I_i=\hbar(n_i+\mu_i/4)$, with $n_i=0,1,\ldots$ and $\mu_i$ the Maslov
index. A multiplet of levels arises as one of the $f$ quantum numbers
runs while all others are fixed. EBK levels of different multiplets have
no reason not to cross as a control parameter $\tau$ is varied. The
exact levels will also cross, provided there are $f-1$ independent
operators commuting mutually and with the Hamiltonian. Assuming such
quantum integrability, we may treat intra-multiplet level dynamics for
$f=1$.

To elucidate the level dynamics for a single-freedom system we may first
look at a non-symmetric double-well potential with a finite barrier
separating the wells; there will be a single eight-shaped separatrix in
the phase space, with the energy of the top of the barrier. For energies
above the ($\tau$-dependent) barrier, all classical orbits lie outside
the separatrix and the function $H(I,\tau)$ is single valued and
monotonic. Variation of $\tau$ cannot bring about crossings of quantum
levels, exact or semiclassical, which stay above the barrier. On the
other hand, for energies below the barrier, separate classical orbits
exist within each loop of the separatrix and for the quantum energy
eigenfunctions are similarly localized. The function
$H(I)$ now has two branches each giving a different action for a given
energy. Neighboring EBK levels whose eigenfunctions live in different
wells can be steered through a crossing by varying $\tau$; only
tunneling corrections turn such crossings into avoided ones. The
closest-approach spacing in an avoided crossing is determined by the
(modulus of the imaginary) action $\Delta I=\int dx\sqrt{2m(E-V(x))}$
across the barrier as $\sim {\rm e}^{-|\Delta I|/\hbar}$; that spacing
may be unresolvably small for $|\Delta I|/\hbar\gg 1$. In the language
of the fictitious gas we could speak of extremely weak repulsion or near
ideal-gas behavior. A few EBK crossings can become, by tunneling,
strongly avoided, with closest approaches of the order of the mean
spacing; these appear for levels near the top of the barrier where
$\Delta I/\hbar$ is of order unity.

\begin{figure}[htb]
\begin{center}
\leavevmode \epsfxsize=0.55\textwidth \epsffile{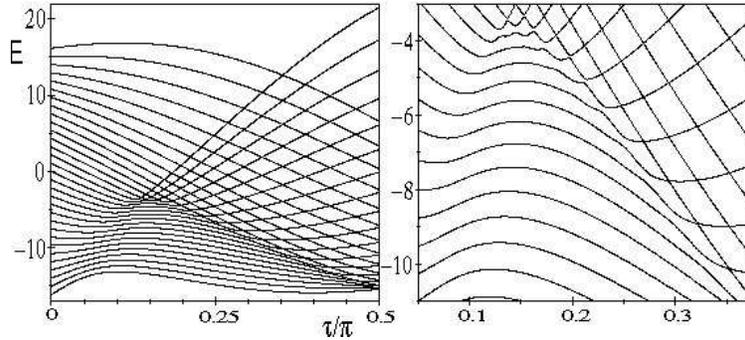}
\caption{Level dynamics of the integrable spin Hamil\-tonian
$H(\tau)\, = (J_{z}+0.4J_{x})\,{\rm cos}\,\tau +(
\frac{4}{2j+1}J_{z}^{2}-0.5J_{z}-\frac{2j+1}{2})\,{\rm sin}\,\tau $ with 
$j=15$.
Strongly avoided crossings are rare exceptions. Right box is a blow-up of left.}
\end{center}
\label{fig.3} 
\end{figure}

Fig.~\ref{fig.3} displays the level dynamics of a single-freedom
system with finite $N$ and illustrates the foregoing discussion. For $\tau$ below a
certain critical value nothing like a crossing occurs since $H(I,\tau)$
remains monotonic and single valued in $I$. For $\tau>\tau_c$ very narrowly avoided
crossings appear which on the scale of the plot look like crossings. In
the neighborhood of a curve $E=E_{\rm sep}(\tau)$ giving the energy of
the separatrix, not drawn in the graph but clearly visible nonetheless,
we see strongly avoided crossings. Inasmuch as a finite fraction of all
$N$ levels can hit the curve $E_{\rm sep}(\tau)$ we may argue that the
number of strongly avoided crossings is linear in $N$ while the number
of unresolved anticrossings is $\sim N^2$. Of course,
the behavior just sketched pertains to Hamiltonians with a single
separatrix. We could easily construct a Hamiltonian with so many
separatrices and consequently so many strongly avoided crossings that
its level dynamics for fixed matrix dimension $N$ would look rather like
what we usually find for chaotic systems; but such a Hamiltonian would at
once be revealed as an impostor by decreasing the effective value of
$\hbar$, i.~e.~going to matrix representations of larger dimension $N$;
the very mechanism making more or less all crossings well avoided for
some sufficiently small $N$ would cast them into negligible minority as
$N\to\infty$ or, equivalently, $\hbar\to 0$.

\if Roughly speaking, therefore, the level dynamics of an integrable system
resembles an ideal gas for sufficiently small $\hbar$. 

We conclude with a remark on chaotic dynamics not faithful to
random-matrix theory. Prominent among these are cases with quantum
localization like the kicked rotator. Two neighboring levels $\varphi_m,
\varphi_{m+1}$ are overwhelmingly likely to pertain to eigenfunctions
without noticeable overlap (in the representation where localization is
manifest). The strength $l_{m,m+1}$ of their repulsion must therefore be
minute as well. Strong repulsion only prevails for groups of levels with
sizably overlapping eigenfunctions. The whole spectrum is thus the
composition of many independent subspectra, similar to the spectra of
integrable systems with $f\geq2$. Again, the PYG is told to behave that
way by its initial data.\fi

We wish to sum up our findings. Under conditions of chaos the mean parametric
distance of avoided level crossings goes to zero in the semiclassical limit
like a power of $\hbar$. Consequently, an average over a classically 
vanishing
control parameter interval reveals universal spectral fluctuations. 
For classically integrable dynamics, on the other hand, level 
repulsion is non-generic.
 
Open to further study is the question as to which, if any, equilibrium
ensembles of the PYG apply to classically integrable dynamics and to chaotic
dynamics not faithful to random-matrix theory, like those with (Anderson)
localization, 
quantum symmetries without classical counterpart
\cite{BogomolnyGeorgeotGiannoniSchmit,KeatingMezzadri} or otherwise
intermediate statistics \cite{Bogomolny}. 
To specify such ensembles we would need a better understanding of 
the (infinity of) constants of the motion of the PYG.

We have enjoyed discussions with Joachim Weber, Yan Fyodorov,
Jon Keating, Hans-J\"urgen St\"ockmann \cite{HS}, and Martin
Zirnbauer, as well as support by the Sonderforschungsbereich ``Unordnung
und Gro{\ss}e Fluktutionen'' der Deutschen Forschungsgemeinschaft 
and Polish KBN Grant No 2 P03B 023 17.


\newpage

\end{document}